\documentclass[aps,prl,showpacs,twocolumn,groupedaddress,amsmath,amssymb]{revtex4}

\usepackage{graphicx}
\usepackage{dcolumn}
\usepackage{bm}
\usepackage{epsfig}

\newcommand\refig[1]{Fig. \ref{#1}}
\begin{document}

\preprint{mecaPRL}

\title{A blue sky catastrophe in double-diffusive convection}

\author{Esteban Meca}
\email[]{esteban@fa.upc.es}
\author{Isabel Mercader}
\author{Oriol Batiste}
\author{Laureano Ram\'{\i}rez-Piscina}
\affiliation{
Departament de F\'{\i}sica Aplicada, Universitat Polit\`ecnica de Catalunya \\
Jordi Girona 1-3, E-08034 Barcelona, Spain}

\date{\today}

\begin{abstract}

A global bifurcation of the blue sky catastrophe type has been found
in a small Prandtl number binary mixture contained
in a laterally heated cavity.
The system has been studied numerically applying
the tools of bifurcation theory.
The catastrophe corresponds to
the destruction of an orbit which, for a large range of Rayleigh numbers, is the
only stable solution. This orbit
is born in a global saddle-loop bifurcation and
becomes chaotic in a period doubling cascade just before its disappearance at the
blue sky catastrophe.

\end{abstract}


\maketitle

Bifurcation theory has long been a very helpful tool in the analysis
of complex dynamics of nonlinear systems
\cite{Kuznetsov,guckenheimer}. Whereas different devised scenarios
have been found in theoretical models with a few variables, there is a
growing interest both in relating real systems with that kind of
models ({\it e.g.} projecting their dynamics to some relevant degrees
of freedom \cite{bhatta}) and in directly analyzing the behavior of
these systems in terms of dynamical systems theory (by studying them
either experimentally or by realistic models). In this context a great
deal of work has been devoted to convection in fluids. Qualitative
changes in the dynamics of fluxes maintained out of equilibrium by
imposed thermal gradients have provided examples of most of the known
bifurcations, and have become a main subject in the area of nonlinear
dynamics.


In this letter we will show the occurrence of a blue sky catastrophe [BSC] in
double diffusive convection. The BSC is a
codimension-1 bifurcation that consists in the destruction of a stable
periodic orbit as its length and period tend to infinity, while the
cycle remains bounded and located at a finite distance from
all the equilibrium solutions \cite{LShilTur00,Kuznetsov}.
This destruction is caused by the collision with a non-hyperbolic
cycle that appears at the bifurcation point.
While approaching the bifurcation the orbit increasingly coils
in the zone where the new cycle will appear, which originates the divergence in
both period and length.
In that point the original cycle becomes an orbit homoclinic to the new cycle.
This type of bifurcation is relatively exotic, but
can easily be found in slow-fast (i.e. singularly
perturbed) systems with at least two fast variables \cite{AShilLShilTur}.


We are interested in double-diffusive fluxes that occur when
convection is driven by simultaneous thermal and concentration
gradients in a binary mixture \cite{Turner85}.  Double-diffusive
convection in cavities with imposed vertical gradients exhibits very
rich dynamics, and has been used as a system to study pattern
formation \cite{cross93} and transition to chaos \cite{KnMoToWe86}.
The case of horizontal gradients, which arises naturally in applications such as crystal growth
\cite{Turner80} or oceanography \cite{Turner85}, has received less attention.
In this work we numerically study this latter configuration for a small Prandtl number
binary mixture. We consider the case when thermal and solutal buoyancy
forces exactly compensate each other, which allows the existence of a
quiescent (conductive) state \cite{GhMo97,XiQeTu97,BaBeKnMo00,BeKn02}.
We have found that in this system there exists a large range of
Rayleigh numbers in which the only stable solution is an orbit that
features a low-frequency spiking behavior.  This orbit appears
associated to a global bifurcation and loses stability when a period doubling
cascade takes place originating a chaotic attractor.
However, the most remarkable feature of this chaotic attractor is its sudden disappearance in
a BSC of the chaotic type.
  As far as we know this is the first example of such bifurcation
in an extended system.

We have considered a binary mixture in a 2-D rectangular cavity of
aspect ratio $\Gamma = d/h = 2$, where $d$ is the length and $h$ is
the height of the cavity. A difference of temperature $\Delta T$ is
maintained between both vertical boundaries. Dimensionless equations
in Boussinesq approximation explicitly read
\begin{eqnarray}
 \partial_t {\bf u} + ({\bf u} \cdot \nabla){\bf u}  &  =
& -\nabla P  + \sigma \nabla^2{\bf u}
\nonumber
\\
 \label{ecs1}
 + \sigma Ra [&\left(1  + S\right)&\left(-0.5 +x/\Gamma\right) + \theta + S C ] \hat{\bf z},
\\
 \label{ecs2}
 \partial_t \theta + ({\bf u} \cdot \nabla) \theta  & =
& - v_x/ \Gamma + \nabla^2 \theta,
\\
 \label{ecs3}
 \partial_t C + ({\bf u} \cdot \nabla) C  & =
& - v_x /\Gamma - \tau \nabla^2
(\theta - C),
\\
 \label{ecs4}
 \nabla \cdot {\bf u}  & =
& 0,
\end{eqnarray}
where ${\bf u} \equiv (v_x,v_z)$ is the velocity field
in $(x,z)$ coordinates, $P$ is the pressure over the density, $\theta$
denotes the departure of the temperature from a linear horizontal
profile. $C$ is the scaled deviation of
the concentration of the heavier component relative to the linear
horizontal profile which equilibrates that of the temperature in the
expression of the mass flux. Lengths and times are scaled with $h$ and
$t_{\kappa}=h^2/\kappa$, respectively, being $\kappa$ the thermal
diffusivity.  The dimensionless parameters are the Prandtl number
${\sigma}=\nu / \kappa$, the Rayleigh number ${ Ra}={\alpha g h^3}
\Delta T / {\nu \kappa}$ and the Lewis number ${\tau} = {D} /
{\kappa}$, where $\nu$ denotes the kinematic viscosity, $g$ the
gravity level, $\alpha$ the thermal expansion coefficient, and $D$ is
the mass diffusivity.  The separation ratio $S=
C_0(1-C_0)\frac{\beta}{\alpha}S_T$ will be taken $S=-1$.  Here, $S_T$
is the Soret coefficient, $C_0$ is the concentration of the heavier
component in the homogeneous mixture, and $\beta$ is the mass
expansion coefficient ($\beta >0$ for the heavier component).

The boundaries are taken to be no-slip and with no mass flux. Lateral
walls are maintained at constant temperatures and at the horizontal
plates a linear profile of temperature between the two prescribed
temperatures is imposed. Thus, boundary conditions are written as
\begin{equation}\label{bc}
{\bf u}={\theta}= {\bf n} \cdot \nabla( C -\theta)=0, \quad \mbox
{at $\partial \Omega$}.
\end{equation}
Notice that these boundary conditions prevent one to absorb the Soret
terms into the equations like in
Refs. \cite{GhMo97,XiQeTu97,BaBeKnMo00,BeKn02}. On the other hand this
system is ${\mathbb Z}_2$-equivariant.  Eqs.  (\ref{ecs1}-\ref{ecs4}),
together with boundary conditions (\ref{bc}), are invariant under a
transformation $\pi$, a central symmetry around the point
$(\Gamma/2,1/2)$, {\it i.e.}
$\pi: (v_x,v_z,\theta,C) \rightarrow (-v_x,-v_z,-\theta,-C), (x,z)
\rightarrow (\Gamma -x,1-z) $.  Hence any solution of these equations
either is $\pi$-invariant (for now on we will call it symmetric) or
its image under $\pi$ is also a solution (constituting a pair of
asymmetric solutions).  This has important consequences on the nature
of its possible bifurcations \cite{Kuznetsov}.

We have obtained time-dependent solutions of
equations (\ref{ecs1}-\ref{ecs4}) and boundary conditions
(\ref{bc}) by using a second order time-splitting algorithm, proposed
in Ref. \cite{HuRa98}, and a pseudo-spectral Chebyshev method for the
space discretization.
Furthermore we have calculated (both stable and unstable) steady solutions and
analyzed their stability by adapting a pseudospectral
first-order time-stepping formulation, as described in Ref.
\cite{MaTu95,BeHeBeTu98,XiQe01}.
The values of the parameters have
been $\sigma=0.00715$ and $\tau=0.03$, close to that characteristic of molten doped
germanium \cite{Kim88,Adornato87}. Spatial discretization has typically been between $ 60
\times 30 $ and $ 90 \times 60 $ mesh grid points.

\begin{figure}
\includegraphics[width=0.35\textwidth,clip]{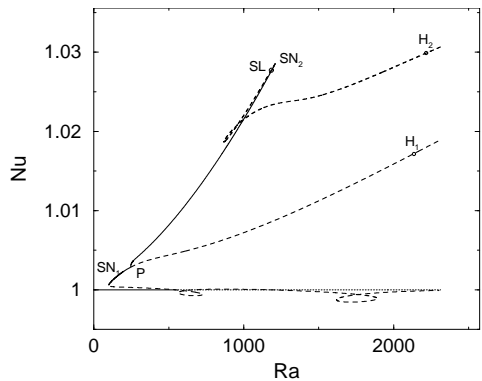}
\caption{Nusselt number of steady solutions versus Ra, and its corresponding
bifurcations. Continuous lines: stable states. Dashed lines: unstable states.}
\label{fig.fig1}
\end{figure}

\begin{figure}
\includegraphics[width=0.45\textwidth,clip]{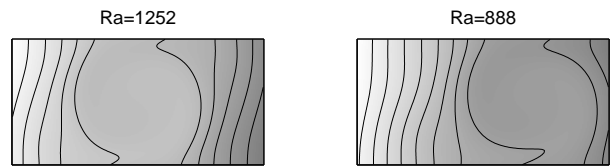}
\caption{Concentration levels of the steady
solutions of the symmetric branch ($Ra=1252$) and the non symmetric
branch ($Ra=888$).}
\label{fig.patterns}
\end{figure}

The scenario provided by the analysis of the
steady solutions is shown in the bifurcations diagram of
\refig{fig.fig1}. In this figure the Nusselt number ({\it Nu}),
defined as the quotient of heat flux through the hot wall with that of
the corresponding conductive solution, is represented for the steady
states as a function of the Rayleigh number ($Ra$). For the sake of
clarity only one asymmetric solution of each pair has been shown. For
small $Ra$ the conductive solution (allowed here by the choice $S=-1$)
is stable, but loses stability, maintaining the symmetry, through a
transcritical bifurcation at $Ra = 541.9$. The supercritical
branch of the bifurcating solution is only stable up to a pitchfork bifurcation
at $Ra=542.4$, following a scenario similar to that
described in Ref. \cite{BaBeKnMo00}.
The interesting behavior in this system originates from the
subcritical branch. This branch gains stability via a saddle node
bifurcation at $Ra=99$ ($SN_{1}$), and loses it again at $Ra=245$ in a
Pitchfork bifurcation ($P$) where a couple of stable asymmetric
branches appear. In \refig{fig.patterns} we represent the concentration for a symmetric (left) and an
asymmetric (right) steady states. We can see that concentration is
roughly homogeneous inside rolls, displacing concentration gradients
to the lateral boundaries.

The asymmetrical steady state is stable until $Ra=1209$, where it loses
stability at a saddle-node bifurcation ($SN_{2}$). The full branch of
asymmetrical steady states is depicted in \refig{fig.fig1}, where we
can see that it changes again the direction at a turning point at
$Ra=865.6$, but without gaining stability. Increasing the Rayleigh number
Hopf bifurcations of the symmetric and asymmetric branches take place
at $H_1$ ($Ra=2137$) and $H_2$ ($Ra=2218$) respectively.  The branch of
symmetric periodic orbits emanating from $H_1$ will play an essential
role in the subsequent evolution of the system.


\begin{figure}
\includegraphics[width=.23\textwidth,clip]{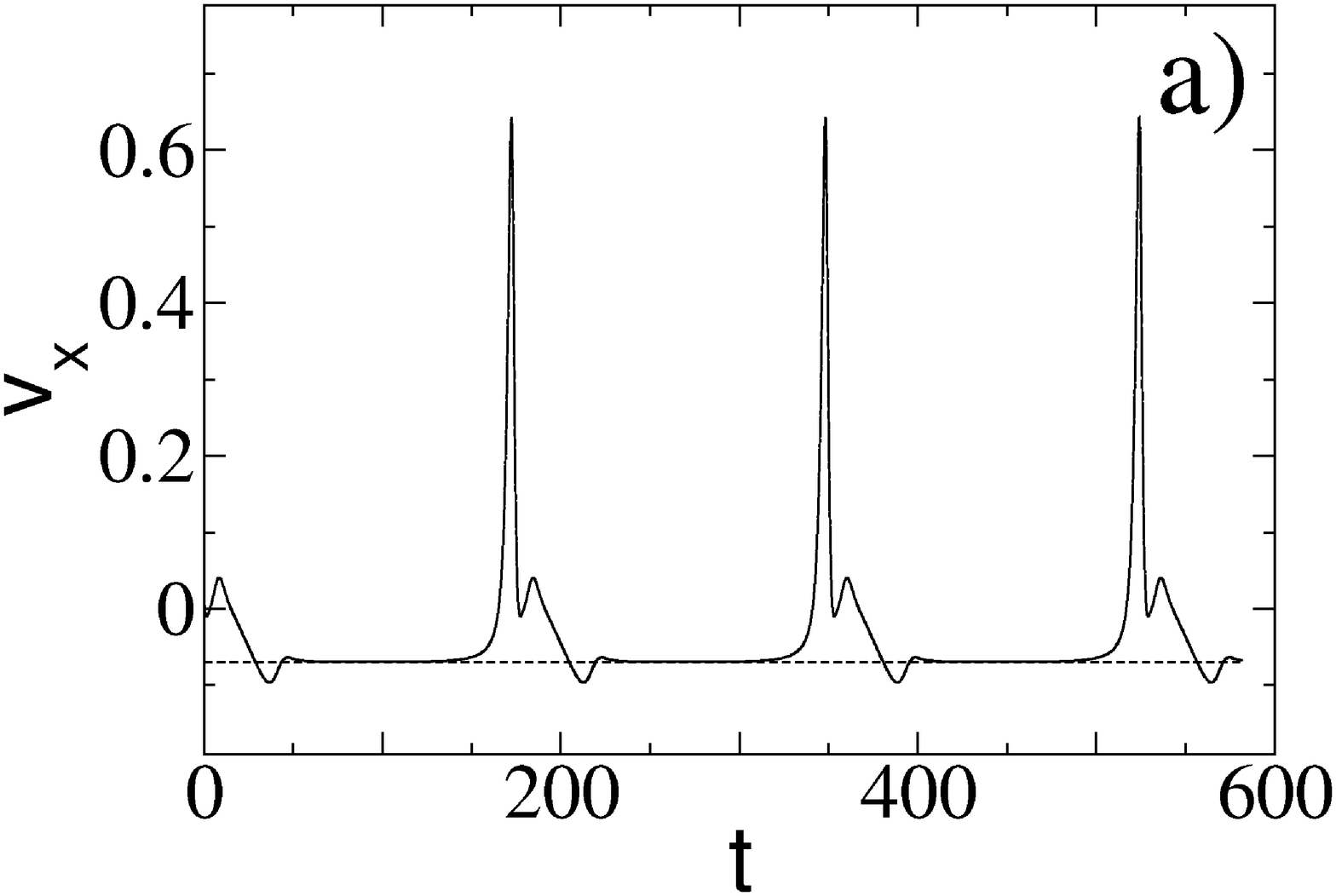}
\hspace{0.0cm}
\includegraphics[width=.23\textwidth,clip]{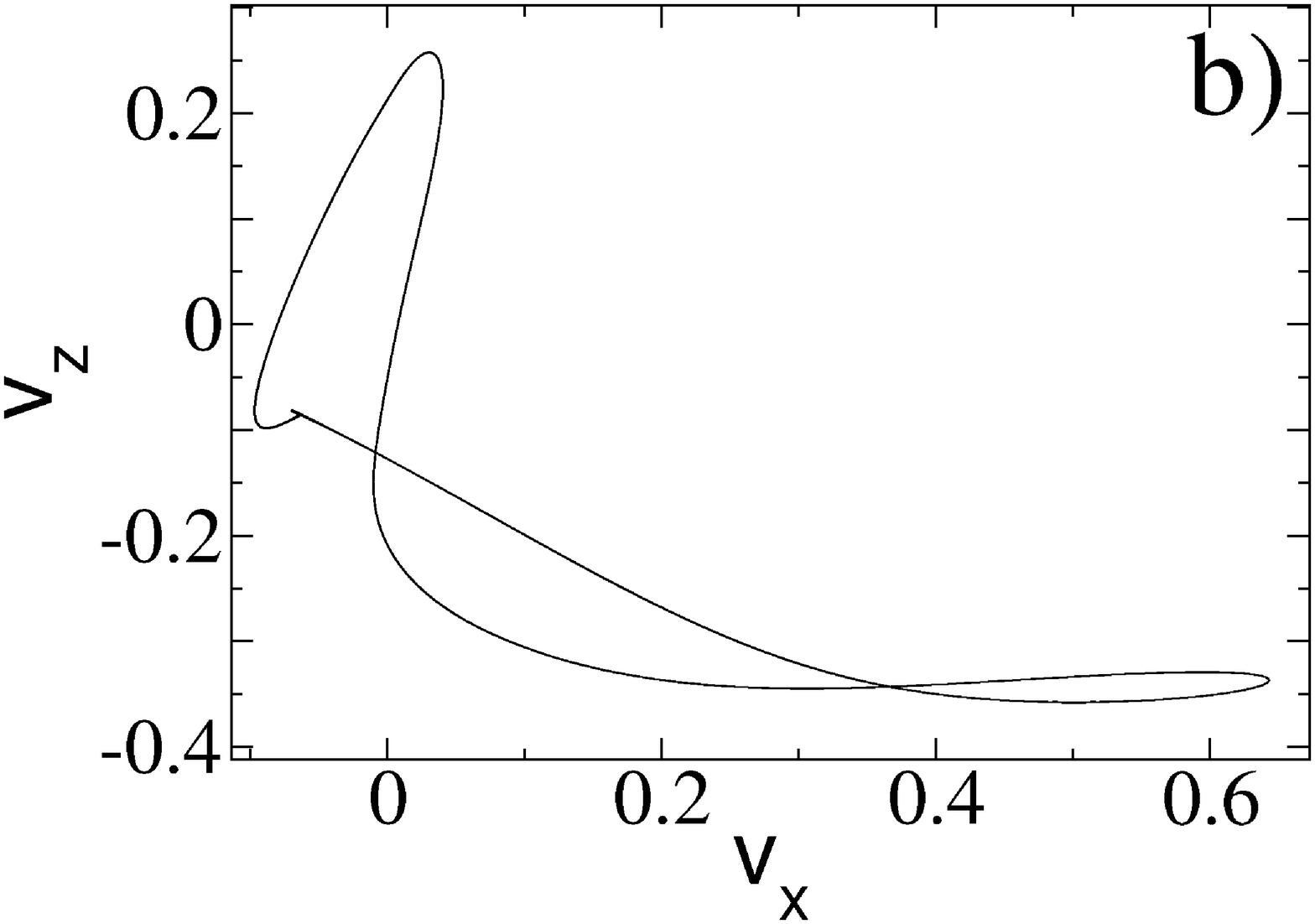}\\

\includegraphics[width=.23\textwidth,clip]{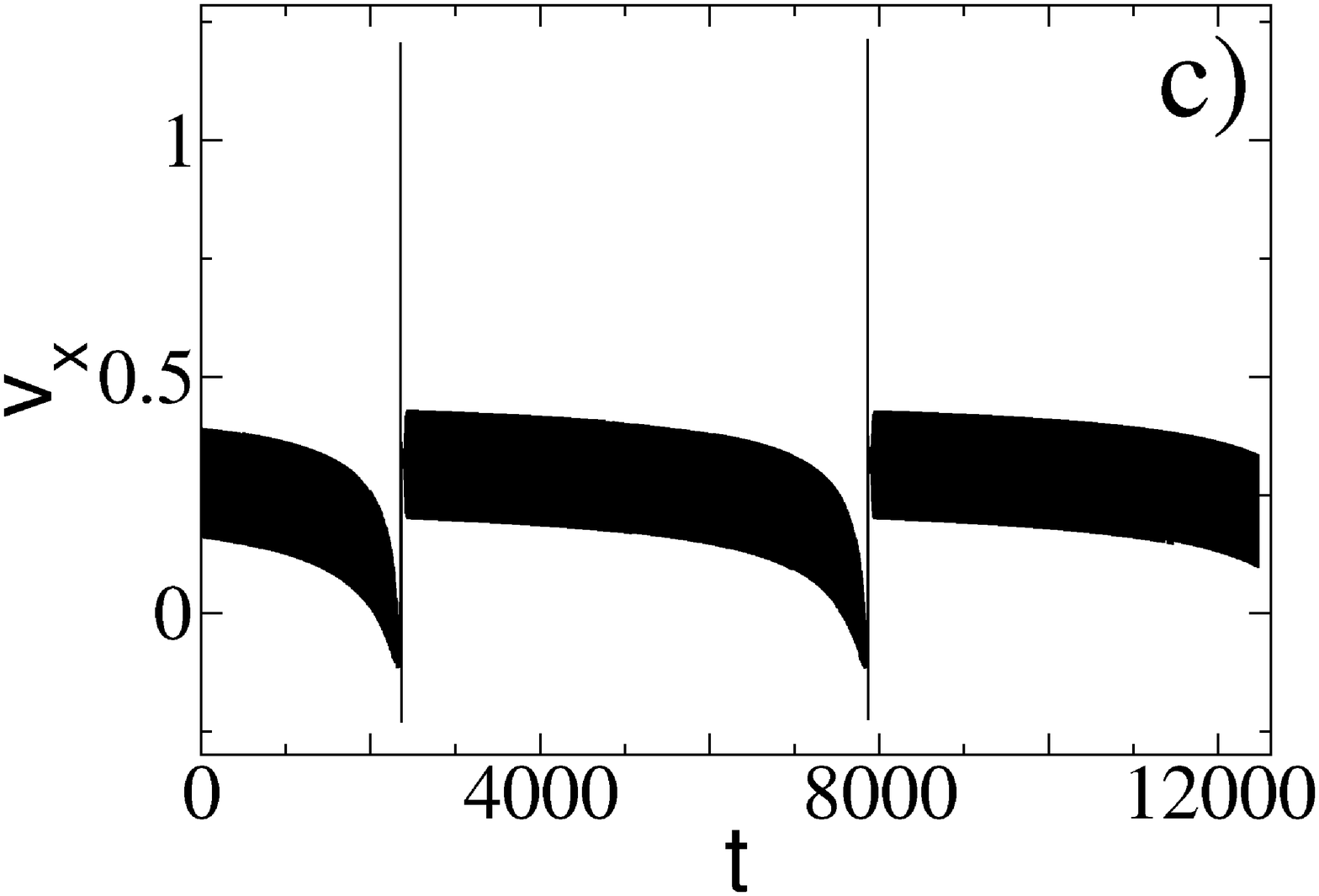}
\hspace{0.0cm}
\includegraphics[width=.23\textwidth,clip]{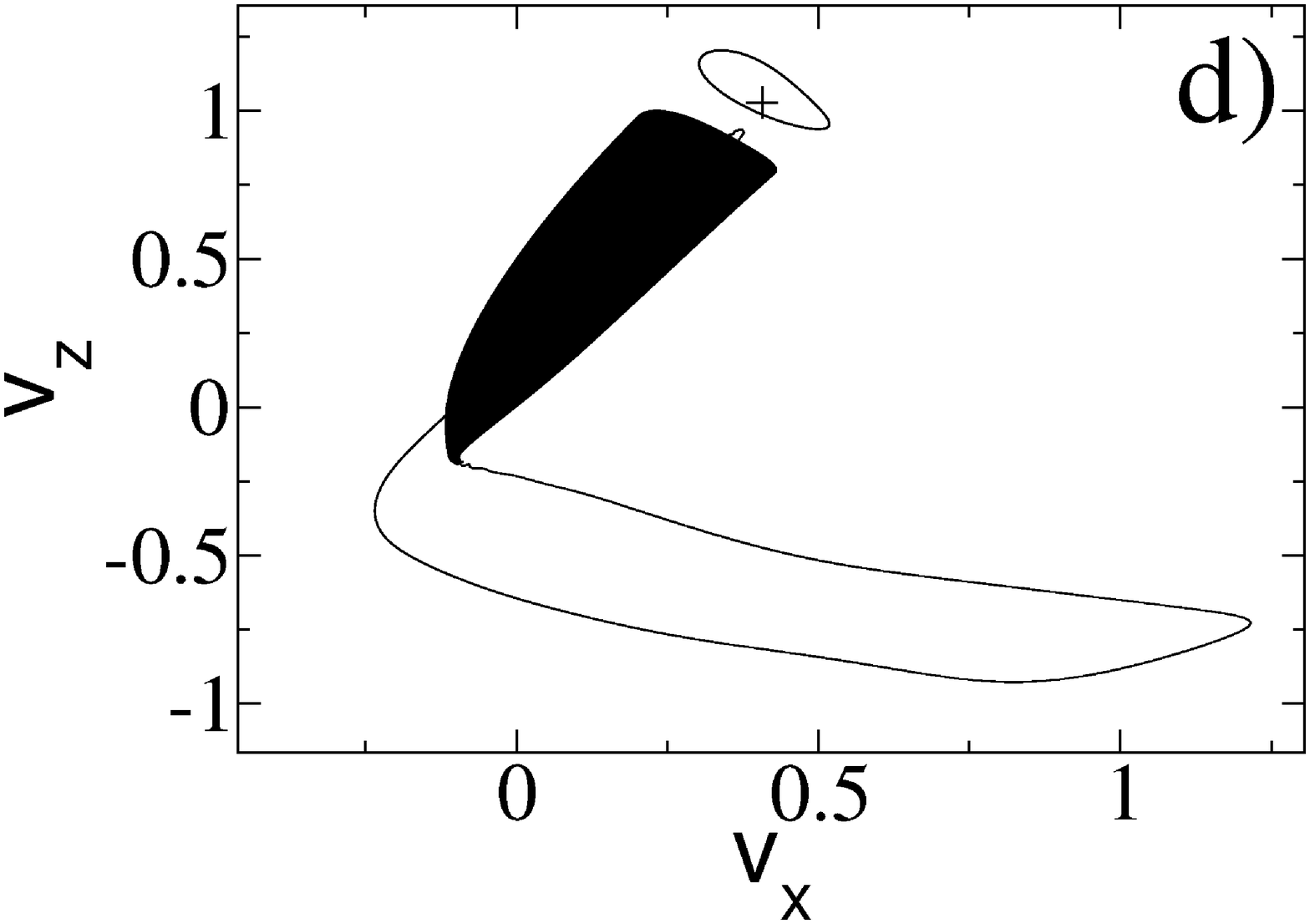}
\caption{ Velocity components of a representative point.
Top: Asymmetric orbit at  $Ra=1183.68$. a) time series, with the value
of the saddle stationary solution marked. b) orbit in the phase space.
Bottom: Attractor at  $Ra=2255$. c) time series. d) attractor
in the phase space with the stable symmetric orbit. The unstable stationary symmetric solution is also shown.}
\label{fig.attract}
\end{figure}

\begin{figure}
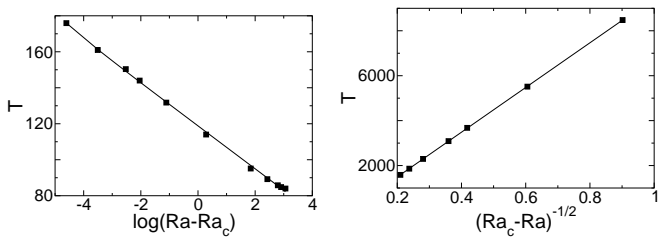

\includegraphics[width=.23\textwidth,clip]{fiteolog_linea.eps}
\hspace{0.1cm}
\includegraphics[width=.235\textwidth,clip]{fiteosqrt_linea_recip.eps}
\caption{Left: logarithmic fit of the periods for the SL connection.
Right: Square-root fit of the periods for the BSC.}
\label{fig.figdivs}
\end{figure}

In the range from $Ra=1209$ until $Ra=2253$ we have found no stable
solution connected with the above branches by local bifurcations.
Integrating the evolution equations we have obtained a branch of
asymmetric periodic solutions that dominates the dynamics of the
system in this range of parameters.  In \refig{fig.attract} we
represent time series and phase space plots of the orbits of this
branch for two different values of the Rayleigh number. The
oscillations first appear in the form of spikes of very large period
(see \refig{fig.attract} a), according to the proximity to a global
saddle-loop bifurcation that occurs at $Ra=1183.67$ (SL) where the orbit
connects with the unstable branch of $SN_{2}$ (see
\refig{fig.fig1}).
The character of this global bifurcation can be inferred from the
logarithmic divergence of the period when the Rayleigh number
decreases toward ($SL$). We have fitted that period to
\begin{equation}
T \sim -\frac{1}{\lambda}\log\left(Ra-Ra_{SL}\right)+A.
\end{equation}
We can see the fit in \refig{fig.figdivs} (left).  The resulting value
 $\lambda{}_{fit}= 0.079 $ results to be quite close to the unstable
eigenvalue $\lambda{}= 0.074$ of the saddle stationary point, as
obtained by the stability calculation.  Near that global
bifurcation the time evolution of the velocity of a representative
point is shown in \refig{fig.attract} (a). The value for the
saddle asymmetric state is also represented. We can see how the
solution spends a long time near it. The spike corresponds
to a rapid and large excursion by the phase space, as seen in
\refig{fig.attract} (b), during which the roll alternately
switches between a confined and a more centered positions (analogous
to the patterns shown in \refig{fig.patterns}).

\begin{figure}
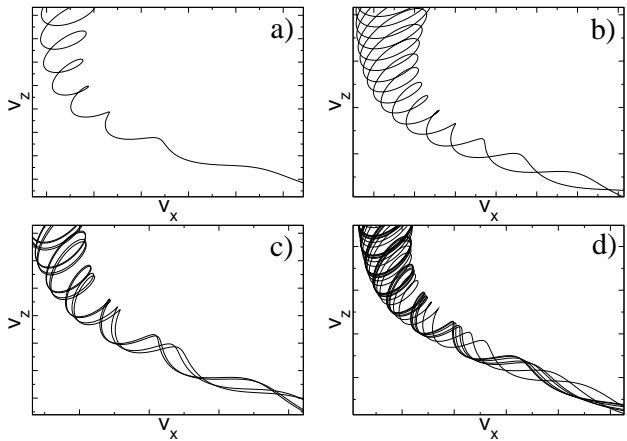

\includegraphics[width=.22\textwidth,clip]{doub11.eps}
\hspace{0.1cm}
\includegraphics[width=.22\textwidth,clip]{doub12.eps}\\
\includegraphics[width=.22\textwidth,clip]{doub13.eps}
\hspace{0.1cm}
\includegraphics[width=.22\textwidth,clip]{doub14.eps}
\caption{Period-doubling cascade (zoom of the tip of the attractor). a) period 1
($Ra=2220$). b) period 2 ($Ra=2232$). c) period 4 ($Ra=2235$). d) chaotic solution ($Ra=2240$).}
\label{fig.pdoub}
\end{figure}

Increasing $Ra$, at $Ra=2137$ the orbit starts to curl, showing ripples
in the time dependence, reflecting the frequency of the unstable
symmetric orbit that appears in $H_1$.  In fact we have
been able to calculate this unstable branch by temporal
evolution forcing the symmetry of the system, and its frequency coincides with that
of the windings of the attractor on all the branch.
If we increase further $Ra$, the asymmetric orbit follows a
period-doubling cascade, becoming chaotic.  This is revealed in the
phase of the winding of the trajectory, as can be seen in
\refig{fig.pdoub} where a detail of the orbit during
the cascade is shown.  This cascade seems to move to slightly higher
$Ra$ values as spatial resolution is increased, but we have not been
able to obtain the precise values due to the extremely large duration
of the orbits in this regime.
 In \refig{fig.attract} (c,d) the attractor thus generated is represented
 at $Ra=2255$. For this value of $Ra$ the symmetric orbit has already
 become stable at a Pitchfork bifurcation ($P_{so}$, at $Ra=2253$), and both
 coexist.
  Very shortly afterwards, the whole
attractor disappears at $Ra = 2257.5$.

This destruction of the attractor exhibits characteristics
that permit to identify it as the chaotic counterpart of the scenario for BSC
bifurcation described in Refs. \cite{LShil97,Kuznetsov}.
Indeed in all the process the attractor remains bounded and at a finite distance
of any steady solution, as required \cite{Kuznetsov}.
The average length and time between spikes
(which are reproducible with variations smaller than 1 over 1000)
diverges as
the windings start to accumulate, which occurs at a specific location in the
attractor.
That indicates that the solution is colliding there with a new cycle
that appears at the bifurcation point, and
to which it becomes homoclinic. Furthermore
this divergence, shown in
\refig{fig.figdivs} (right), is very well fitted by a square-root law:
\begin{equation}
T\sim \frac{A}{\sqrt{Ra_c-Ra}}+B.
\end{equation}
This law of divergence particularly corresponds to that scenario, since
it demonstrates that the new cycle to which
the attractor is connecting is the saddle-node of two orbits ($SNO_1$).
In principle there are several
possibilities for the topology of the attractor \cite{LShilTur00}.
In our case the successive windings are braided by the tip of the attractor into
an almost one dimensional tube or filament (\refig{fig.pdoub}). This
filament reintroduces the orbit into the vicinity of the saddle-node orbit,
and it starts winding again accumulating curls near it. Therefore, in the limit,
the attractor has the topology of a French Horn. This feature is also shared with
Ref. \cite{LShil97}.

After the BSC, one would expect the system to reach the stable member of the pair
of asymmetric solutions born at $SNO_1$. On the contrary,
simulations show that the system evolves through a extremely long transient,
during which the trajectory accumulates curls near the
saddle-node before being rejected to the symmetric orbit that became stable at
$P_{SO}$. That could mean either that the stability range of the asymmetric orbit
is very small (which would require a much  finer exploration in $Ra$ to find it,
a formidable task in this slow regime), or that its basin of attraction is very
reduced (and the nearby symmetric orbit attracted all the calculated orbits).

\begin{figure}
\includegraphics[width=.31\textwidth]{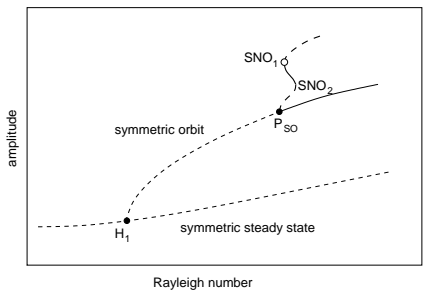}
\caption{Diagram of the conjectured unstable asymmetric orbit (thick line)
and its connections to other branches}
\label{fig.con}
\end{figure}

We propose that the $SNO_1$ is located in the branch of
unstable asymmetric orbits created at the pitchfork bifurcation where
the new orbit becomes stable ($P_{so}$). This hypothetical scenario is shown in
\refig{fig.con}, and is the simplest one in which the attractor presents at the
BSC an homoclinic connection to a branch coming from known solutions. This
conjecture requires the unstable asymmetric branch to gain stability in a first
saddle node bifurcation $SNO_2$ and to lose it again at the $SNO_1$, as can be
seen in \refig{fig.con}. The coincidence of the frequency value of the symmetric
orbit at $P_{SO}$, $\omega_{SO}=7.01$, to that of the windings of the attractor,
$\omega=7.01$, is consistent with such connection. A small distance between $SNO_1$
and $SNO_2$ would explain the reduced stability domain of the asymmetrical solution.
 Finally, the presence of the additional saddle node orbit $SNO_2$ in the
proximity would slow down the dynamics for $Ra$ slightly above, making the
transient to the symmetric solution very long, as it is actually observed.

One could devise more complex scenarios for the occurrence of this BSC. For example,
the attractor could be destroyed on a boundary crisis associated to global connections of
the unstable orbits coming from the period doubling bifurcations.


Finally, it is worth remarking than
the BSC displayed by this system is robust against small changes in the value of
the separation ratio $S$. In particular we have obtained a similar BSC
in simulations performed with $S=-0.99$. That means that the
additional symmetry introduced in the system by the special value
$S=-1$ is not an essential ingredient of the phenomena described here.

This work was financially supported by
Direcci\'on General de Investigaci\'on Cient\'{\i}fica y T\'ecnica
(Spain) (Projects BFM2003-00657 and BFM2003-07850-C03-02) and
Comissionat per a Universitats i Recerca (Spain) Projects
(2001/SGR/00221 and 2002/XT/00010).  We also acknowledge computing
support from CEPBA (Spain).
E.M. acknowledges a grant from Ministerio de Educaci\'{o}n, Cultura y Deporte (Spain).

\bibliographystyle{apsrev}
\bibliography{mecaprl2}


\end{document}